\documentclass{elsart}
\usepackage{graphicx,amssymb}
\journal{Chemical Physics Letters}
\begin{document}

\begin{frontmatter}

\title{A fully polarizable and dissociable potential for water}

\author[dft]{E. Lussetti}
\author[dft,demo]{G. Pastore} and
\author[dft,demo]{E. Smargiassi\corauthref{cor}}
\corauth[cor]{Corresponding author}
\ead{smargiassi@ts.infn.it}

\address[dft]{Department of Theoretical Physics, University of
Trieste, Strada Costiera 11, I-34014 Trieste (Italy)}
\address[demo]{INFM-DEMOCRITOS, National Simulation Center}

\begin{abstract}

A new classical interaction potential for water simulations is
presented. Water is modeled as a fully dissociable set of atoms with a
point dipole, determined self-consistently, on every oxygen atom. The
oxygen polarizability is not fixed but depends on the geometry of the
system. We show that, in spite of the limited number of free
parameters, the model reproduces the geometrical and vibrational
properties of microclusters in a satisfactory way.

\end{abstract}
\end{frontmatter}

Water, being arguably the most important liquid in our life, has
always attracted a great deal of attention from physicists and
chemists. It is also a rather difficult substance to model, and, in
spite of vast improvements in numerical techniques, {\sl ab-initio}
simulations are not yet able to simulate more than a hundred molecules
for a reasonable time. Many classical potentials have been proposed,
but very few of them are able to take into account all the important
structural and dynamical features of water. Dissociability, for
instance, is important in order to describe bulk water and its
hydration and proton-transfer properties. Polarizability is crucial in
the description of structural properties.

In this Letter we propose a new dissociable and polarizable potential
for water and validate it by means of the study of the geometrical and
vibrational properties of the water molecule and of the water dimer
and trimer; results on larger clusters, and a full comparison with
other potentials, will be presented in a full paper. The validation
process we used is especially stringent as we predict a range of
physical properties much larger than usually done in this kind of
studies. We believe this gives a better guarantee of transferability
to our potential than the usual procedures. Our potential contains a
small number of parameters and is based on reasonable physical
assumptions, which include some quantum effects previously
neglected. Its aim is to reunite the advantages of dissociable and
polarizable potentials while at the same time keeping a comparatively
simple and physically transparent form.

We assume that, in the molecule, most of the electric charge of each
hydrogen is transferred to the oxygen. We also assume that the
electronic polarizability of the molecule is entirely due to the
oxygen. Several models already exist that allow one to treat
successfully ionic crystal structures (see for instance
\cite{havinga67,galli84,wilson96}), and we will start from there. The
potential we propose has the following form:
$$U=U_{ZZ}+U_{BM}+U_{DD}+U_{VW}+U_{ZD}+U_{SR}.$$
In the term
\begin{equation}\label{Uzz}
U_{ZZ}=\frac{1}{2}\sum_{i\neq{j}}\frac{Z_{i}Z_{j}}{r_{ij}}
\end{equation}
$Z_o$ and $Z_h$ indicate ionic charges and are considered as
adjustable parameter, bound by the charge neutrality constraint
$Z_o+2Z_h=0$ and \( \textbf{r}_{ij}=\textbf{r}_i-\textbf{r}_j,\) where
\textbf{r}$_i$ is the position of the i-th nucleus. The Born-Mayer
repulsive term
\begin{equation}\label{Ubm}
U_{BM}=\frac{1}{2}\sum_{i\neq{j}}f(\rho_i+\rho_j)exp\left(\frac{R_i+R_j-r_{ij}}{\rho_i+\rho_j}\right)
\end{equation}
includes the adjustable parameters $f, R_o, R_h, \rho_o,\rho_h$. The
term
\begin{equation}\label{Udd}
U_{DD}=\frac{1}{2}\sum_{i\in{O}}\,\sum_{j\in{O},j\neq{i}}\left(\frac{\textbf{p}_i\cdot\textbf{p}_i}{r^{3}_{ij}}-\frac{3(\textbf{p}_i\cdot\textbf{r}_{ij})(\textbf{p}_j\cdot\textbf{r}_{ij})}{r^{5}_{ij}}\right)+\sum_{j\in{O}}\frac{p^{2}_j}{2\alpha_o}
\end{equation}
describes dipole-dipole interaction, \(\textbf{p}_i\) being a point
dipole located on the $i-$th oxygen, and the bare oxygen
polarizability $\alpha_o$ is taken as a further parameter; since
hydrogen is almost entirely deprived of electrons, it is considered
unpolarizable. The Van der Waals term is written as
\begin{equation}\label{Uvw}
U_{VW}=-\frac{1}{2}\sum_{i}\,\sum_{j\neq{i}}\frac{c_i c_j}{r^6_{ij}}
\end{equation}
while the term
\begin{equation}\label{Uzd}
U_{ZD}=\sum_{j\in{O}}\sum_{i,i\neq{j}}\textbf{p}_j\cdot{Z_{i}}\frac{\textbf{r}_{ij}}{r^3_{ij}}
\end{equation}
describes the charge-dipole interaction. The term
\begin{equation}\label{Usr}
U_{SR}=-\sum_{i\in{H}}\sum_{j\in{O}}\textbf{p}_j\cdot{B}(r_{ij})\frac{\textbf{r}_{ij}}{r_{ij}}, \hspace{1em}
B(r_{ij})=\frac{F}{\alpha_o}f\,
exp(\frac{R_{sr_i}+R_{sr_j}-r_{ij}}{\rho_i+\rho_j})
\end{equation}
is taken from Ref.\cite{galli84}, where the polarizability in the
shell model is studied. Intuitively, this term reduces the
polarizability of oxygen by taking into account the interaction
between its electronic charge and that of the closest
hydrogen atoms, so it is not surprising that it has the same form of
Eq. \ref{Ubm}. In principle the values considered for
$R_{sr_o}+R_{sr_h}$ (only the sum of these parameters is relevant) are
independent of $R_o+R_h$, but in practice they are only slightly
different, and in some previous studies \cite{galli84} they were set
as equal.

In almost all this work we set $\rho_h=0$, that is we neglected the
short-range hydrogen-hydrogen interactions given by Eqs. \ref{Ubm} and
\ref{Usr}. In short, our potential is characterized by 11 independent
parameters, which are listed in table \ref{table_par} together with
their optimized values.

\begin{table}[h]
\begin{tabular}{|ll|}
\hline
$Z_h=0.80e$ & $\rho_h=0$ \\
$Z_o= -2 Z_h =-1.60e$ & $\rho_o=0.209\AA$\\
$f=0.03405e^2\AA^{-2}$ & $\alpha_o=1.027\AA^3$\\
$R_h=0.012\AA$ & $F=0.29e^{-1}\AA^3$\\
$R_o=1.718\AA$ & $c_h=0.01e\AA^{5/2}$\\
$R_{sr_h}+R_{sr_o}=1.77\AA$ & $c_o=3.9e\AA^{5/2}$ \\
\hline
\end{tabular}
\caption{Optimized set of parameters}
\label{table_par}
\end{table}

It is worth noting that the optimized value of $\alpha_o$ lies
approximately midway between the polarizabilities of oxygen and of the
water molecule. We fitted several physical quantities we considered of
primary interest: of the monomer, the geometrical data, the
vibrational frequencies and the dipole moment, all of them known
experimentally.  Of the clusters, the geometrical data relative to the
equilibrium configurations, which are known experimentally and
sometimes from {\sl ab-initio} calculations. The equilibrium
geometries were found using a simple steepest-descent algorithm, with
$\textbf{p}_i$ determined self-consistently, while the vibrational
frequencies were obtained via numerical calculation and
diagonalization of the dynamical matrix. After fixing the parameters
of the potential, as a further test we calculated the cluster binding
energies, the vibrational frequencies and some configurations
considered as local minima of the potential energy as reported also in
some \textit{ab-initio} studies, comparing all these data with those
found in the literature.

In the rest of this paper we shall use the following symbols: $\mu $
for the dipole moment of the molecule; $\nu_{ss}$, $\nu_{as}$ and
$\nu_b$ for the molecule normal modes, that is, respectively,
\textit{symmetric stretching}, \textit{asymmetric stretching} and
\textit{bending}; $\overline{\mu}(molec)$ is the average molecular
dipole moment in the clusters.The subscripts $b$ and $f$ indicate the
bond and free hydrogens, respectively.

\begin{table}[h]
\begin{tabular}{|l|l|l|l|}
\hline
 &\textit{Exptl data} & \textit{ab-initio}\cite{xantheas93}& \textit{This work}  \\
\hline
d(O-H) & 0.957\AA\cite{benedict56} & 0.965\AA &0.961\AA \\
$\theta_{H-O-H}$ & 104.5$^o$\cite{benedict56}& 103.8$^o$ & 104.5$^o$ \\
$\nu_b$& 1595$cm^{-1}$\cite{kuchitsu65} & 1623$cm^{-1}$&1158$cm^{-1}$ \\
$\nu_{ss}$ & 3657$cm^{-1}$\cite{kuchitsu65}& 3807$cm^{-1}$ & 3609$cm^{-1}$ \\
$\nu_{as}$ & 3756$cm^{-1}$\cite{kuchitsu65}& 3936$cm^{-1}$ & 3234$cm^{-1}$\\
$\mu$ & 1.86D\cite{lovas78}  & 1.87D\cite{gregory97}&2.07D \\
\hline
\end{tabular}
\caption{Molecule data}
\label{tab_mol}
\end{table}

The results for the molecule are shown in table \ref{tab_mol}. As one
can see, the overall agreement with the experimental data is good, the
only quantity in error of more than 15\% being $\nu_b$. The
frequencies $\nu_{ss}$ and $\nu_{as}$ are in the wrong order, a
feature common in our potential, but given the small difference
between them this does not look too worrying. Attempts made to fix the
discrepancy led to an oxygen-oxygen repulsion that is too small.

The numerical results on the dimer are collected in table
\ref{tab_dim1}. The ground-state geometry is, qualitatively, the usual
one; $\theta_d$ is the angle between the bisector of the H-O-H angle
of the donor molecule and the O-O direction, $\theta_a$ is the same
for the acceptor. $D_e$ is the dimer dissociation energy.

\begin{table}[h]
\begin{tabular}{|l|l|l|}
\hline &\textit{Exptl}$^*$/\textit{ab-initio}$^\dag$ \textit{data}&\textit{This work} \\
\hline $d(O-O)$ &2.98\AA\cite{dyke77}$^*$ & 3.20\AA \\
$\theta (O-H_b-O)$ &$174^o\pm 10^o$\cite{dyke77}$^*$ & 171$^o$ \\
$d(O-H)_b$ & 0.972\AA\cite{xantheas93}$^\dag$ & 0.98\AA \\
$d(O-H)_f$ & 0.964-0.966\AA\cite{xantheas93}$^\dag$ & 0.960-0.968\AA \\
$\theta_a$ &$57^o\pm 10^o$\cite{odutola80}$^*$ & 92$^o$ \\
$\theta_d$ &$51^o\pm 10^o$\cite{odutola80}$^*$ & 57$^o$ \\
$D_e$& 0.23eV\cite{dyke77}$^*$ & 0.16eV \\
$\overline{\mu}(molec)$ & 2.10D\cite{gregory97}$^\dag$ & 2.22D \\
\hline
\end{tabular}
\caption{Dimer structural data}
\label{tab_dim1}
\end{table}

We found that, as $c_o$ and $R_o$ vary, when the oxygen-oxygen
separation increases then $\alpha_o$ decrease by about the same
amount, and vice versa: this is clearly an electrostatic effect due to
the attraction of the hydrogens in the acceptor molecule and the
oxygen of the donor molecule. In addition, the angle $\theta_a$
increases, as one would expect, as the molecular dipole moments
decrease and tend not to align. The excessive value of $\theta_a$ is
therefore a price we have to pay in order to obtain a dipole moment
not too high. We also verified that, as $d(O-O)$ increases, $D_e$
decreases. It is thus not surprising that, since $d(O-O)$ is
overestimated, $D_e$ is underestimated.

\begin{table}[h]
\begin{tabular}{|l|l|l|l|}
\hline
 & \textit{Exptl data} & \textit{Ab-initio} (MP2)    & \textit{This work}\\
\hline
\textit{Intermolecular} & \cite{bentwood80} & \cite{xantheas93}
 \hspace{4mm} \cite{frisch86}   &  \\
$\omega_{12}$: Donor torsion  & -  &141 \hspace{4mm}  155    & 107  \\
$\omega_{11}$: Acceptor twist & - &147  \hspace{4mm}  193  & 238  \\
$\omega_8$: Acceptor bend   & 243 &155 \hspace{4mm}   178  & 262  \\
$\omega_7$: O-O stretch  & 155 &185  \hspace{4mm}  220  & 94  \\
$\omega_6$: In-plane donor wag   & 320 & 342  \hspace{4mm}  398 & 487  \\
$\omega_{10}$: Out of plane donor bend & 520 & 632 \hspace{4mm}   715 & 722 \\
 &  &  &   \\
\textit{Intramolecular}  & \cite{fredin77} &  &   \\
H-O-H acceptor bend & 1601& 1624 \hspace{2mm} 1615 & 1196 \\
H-O-H donor bend & 1619 & 1651  \hspace{2mm}1636& 1303 \\
Donor bridge O-H stretch & 3530 & 3712  \hspace{2mm}3539 &3127 \\
Acceptor sym. O-H stretch & 3600 & 3799  \hspace{2mm}3659 &3590\\
Donor free O-H stretch & 3730 & 3907  \hspace{2mm}3719 &3550 \\
Acceptor asym. O-H stretch & 3745 & 3929  \hspace{2mm}3746  &3215 \\
\hline
\end{tabular}
\caption{Dimer normal modes (in $cm^{-1}$)}
\label{tab_dim2}
\end{table}

Result for the dimer normal modes are shown in table
\ref{tab_dim2}. For their definition we follow Ref.\cite{xantheas93}
and their identification Refs
\cite{bentwood80,reimers84,frisch86}. The ordering of the
intermolecular frequencies we find is not in complete agreement with
experimental or \textit{ab initio} data; it is the same found by
\cite{xantheas93}, with the exception of the ``O-O stretch'' mode. The
experimental data show that the intramolecular frequencies are not
much different from those of the monomer; the main difference consists
of an increase of the bending frequencies and a decrease of the
symmetric and asymmetric stretching frequencies. These trends,
confirmed by \textit{ab initio} calculations (\cite{xantheas93} and
\cite{estrin96}), are well reproduced in our results.

\begin{figure}
\includegraphics[height=6cm,width=1.\textwidth]{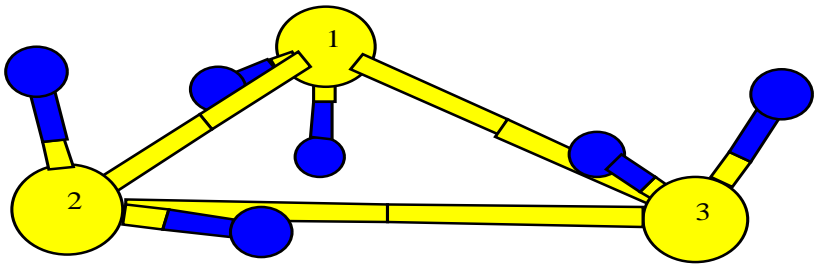}
\caption{Trimer equilibrium geometry}
\label{trimero}
\end{figure}

\begin{table}[h]
\begin{tabular}{|l|l|l|}
\hline
 & \textit{Exp. data}$^a$ & \textit{This work}\\
\hline
d(O-O) & 2.94-2.97-2.97\AA & 3.10-3.09-3.25\AA  \\
$\theta(O_1-H_b-O_2)$ & 152 & 163 \\
$\theta(O_2-H_b-O_3)$ & 150 & 156\\
$\theta(O_3-H_b-O_1)$ & 153 & 164\\
$H_{b1}-(O_1-O_2-O_3)$& 180 & 175\\
$H_{b2}-(O_2-O_3-O_1)$& 180 & 180\\
$H_{b3}-(O_3-O_1-O_2)$& 180 & 181\\
\hline
 & \textit{ab initio} & \textit{This work} \\
\hline
$d(O-H)_b$ &0.977\AA$^e$  &  0.99-1.00\AA \\
$d(O-H)_f$ & 0.964\AA$^e$ &  0.963-0.965-0.966\AA \\
$H_{f1}-(O_1-O_2-O_3)$& 118$^e$ & 93\\
$H_{f2}-(O_2-O_3-O_1)$& 237$^e$ & 268\\
$H_{f3}-(O_3-O_2-O_1)$& 231$^e$ & 256\\
$D_e/N_{mol}$ & 0.22eV$^c$ & 0.15eV\\
$\overline{\mu}(molec)$ &  2.31D$^d$  & 2.37D   \\
Frequency range ($cm^{-1}$) &   &  \\
\textit{intermolecular} & 101-730  $^b$ &  75-1083 \\
                         & 158-863 $^e$    &   \\
                         & 185-951 $^f$    &   \\
\textit{intramolecular} & 1753-4215 $^b$ &  1288-3544 \\
                         & 1632-3898 $^e$ &  \\
                         & 1623-3727 $^f$ &  \\
\hline
\end{tabular}
\caption{Trimer data
$^a$Ref.\cite{campbell85}.$^b$Ref.\cite{xantheas93},
HF.$^c$Ref.\cite{laasonen93},
LDA+GC.$^d$Ref.\cite{gregory97}.$^e$Ref.\cite{xantheas93},
MP2.$^f$Ref.\cite{estrin96}, LDA+GC.}
\label{tab_trim}
\end{table}

\begin{figure}
\includegraphics[height=0.3\textheight,width=0.5\textwidth]{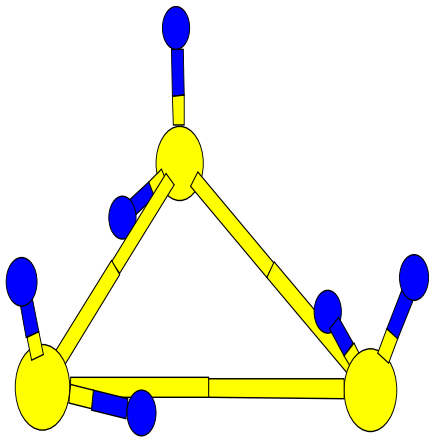}
\caption{Trimer: the lowest metastable state}
\label{mintrim33ps}
\end{figure}

\begin{figure}
\includegraphics[height=0.3\textheight,width=0.5\textwidth]{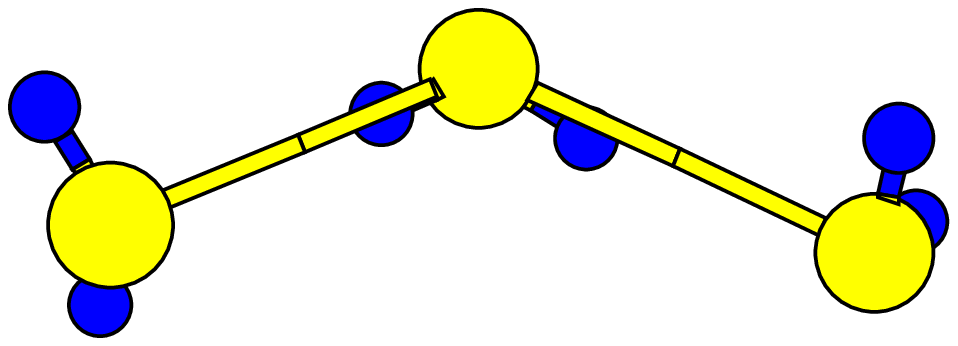}
\caption{Trimer: the next-lowest metastable state}
\label{mintrim32bisps}
\end{figure}

The equilibrium geometry of the trimer is shown in fig.\ref{trimero}
and the numerical data are collected in table \ref{tab_trim}. The
longest side of the triangle corresponds to the smallest H-bond
angle. This is characteristic of our potential. We obtained two
relative minima, shown in fig. \ref{mintrim33ps} and
\ref{mintrim32bisps}. The first is 0.06 eV above the energy in the
absolute minimum, the second 0.15 eV. Their ordering in energy is the
same as in the \textit{ab initio} calculations \cite{klopper95,mo92},
respectively 0.04 and 0.24 eV.

As an illustration of the overall quality of our potential, in Table
\ref{trends} we show the trends of two important quantities --- the
average oxygen-oxygen distance and the average molecular dipole moment
--- with increasing number $n$ of molecules in the cluster. To make
trends clearer, for $n<8$ the lowest-energy cyclic clusters have been
considered, even when they are not the absolute energy minima. It is
clear that the trends from {\it ab initio} calculations are well
reproduced and the absolute errors actually decrease with $n$.

\begin{table}[h]
\begin{tabular}{|l|l|l|l|l|l|l|}
\hline
 $n$& 1 & 2 & 3 & 4 & 6 & 8 \\
\hline
$\overline {d(O-O)}(\AA)$ This work & - & 3.20   & 3.14  & 2.80  &  2.77 &    \\
$\overline {d(O-O)}(\AA)$ \cite{xantheas93} & - & 3.03 & 2.93 & 2.88 & 2.86 &\\
$\overline \mu(molec)(D)$ This work& 2.07 & 2.22 & 2.37 & 2.76 & 2.82 & 2.77 \\
$\overline \mu(molec)(D)$ \cite{gregory97} & 1.86 & 2.10 & 2.31 & 2.55 & 2.70& 2.72   \\
\hline
\end{tabular}
\caption{Trends with varying $n$ in cyclic clusters and the octamer,
  compared with {\it ab initio} results}
\label{trends}
\end{table}

In conclusion, we have shown that the model reproduces the basic
physical properties of microclusters in a satisfactory way. In
particular, we are able to obtain bond lengths and dipole moments
within 10-20\% of experimental and ab-initio values, and to reproduce
qualitatively not only all the ground-state geometries, but also
several local minima. This remains true when the potential is applied
to larger clusters, as we will show in a forthcoming paper. It should
be stressed that for almost none of the potentials found in the
literature one can find a comparison with experimental or {\sl
ab-initio} data as extensive and thorough as ours. Moreover, the
limited number of parameters we used makes it more likely that their
values are close to the real ones. We therefore believe that our
potential is a promising one with regard to transferability and
accuracy.

This work has been funded by a MIUR-COFIN grant.

\end{document}